\documentclass[twocolumn,aps,tightenlinefs,superscriptaddress,preprintnumbers]{revtex4}
\usepackage{graphicx,color,dcolumn,booktabs,bm}
\usepackage{longtable,lscape}
\usepackage{amsmath}
\usepackage{indentfirst}
\usepackage{epsfig}
\usepackage{feynmf}   
\usepackage{epstopdf}   
\usepackage{slashed}  
\usepackage{cases}
\usepackage{color}
\usepackage{multirow}
\usepackage{ulem}
\usepackage{graphicx,color,dcolumn,booktabs,bm}
\usepackage{verbatim}
\usepackage[colorlinks,linkcolor=red,anchorcolor=green,citecolor=blue]{hyperref}
\usepackage{mathrsfs}
\usepackage{rotating}
\usepackage{threeparttable}
\usepackage{lineno}
\usepackage{verbatim}
\usepackage{orcidlink}

\newcommand{\BR}{{\cal B}}

\newcommand{\beq}{\begin{equation}}
\newcommand{\eeq}{\end{equation}}
\newcommand{\bitm}{\begin{itemize}}
\newcommand{\eitm}{\end{itemize}}

\newcommand{\ones}{\Upsilon(1S)}
\newcommand{\twos}{\Upsilon(2S)}

\newcommand{\threes}{\Upsilon(3S)}
\newcommand{\fours}{\Upsilon(4S)}
\newcommand{\fives}{\Upsilon(5S)}

\newcommand{\tabincell}[2]{\begin{tabular}{@{}#1@{}}#2\end{tabular}}

\begin{document}
\hyphenpenalty=10000

\title{\quad\\[0.1cm]\boldmath  Measurement of the branching fraction of $\Xi_{c}^{0}\to \Lambda_{c}^{+}\pi^{-}$ at Belle}

\noaffiliation
\author{S.~S.~Tang\,\orcidlink{0000-0001-6564-0445}} 
\author{Y.~B.~Li\,\orcidlink{0000-0002-9909-2851}} 
\author{C.~P.~Shen\,\orcidlink{0000-0002-9012-4618}} 
  \author{I.~Adachi\,\orcidlink{0000-0003-2287-0173}} 
\author{H.~Aihara\,\orcidlink{0000-0002-1907-5964}} 
\author{D.~M.~Asner\,\orcidlink{0000-0002-1586-5790}} 
\author{H.~Atmacan\,\orcidlink{0000-0003-2435-501X}} 
\author{T.~Aushev\,\orcidlink{0000-0002-6347-7055}} 
\author{R.~Ayad\,\orcidlink{0000-0003-3466-9290}} 
\author{V.~Babu\,\orcidlink{0000-0003-0419-6912}} 
% ...
\collaboration{The Belle Collaboration}

\begin{abstract}

Based on a data sample of 983 fb$^{-1}$ collected with the Belle detector at the KEKB asymmetric-energy $e^+e^-$ collider, we present the study of the heavy-flavor-conserving decay $\Xi_{c}^{0}\to \Lambda_{c}^{+}\pi^{-}$ with $\Lambda_{c}^{+}$ reconstructed via its $pK^{-} \pi^{+}$ decay mode. The branching fraction ratio $\mathcal{B}(\Xi_{c}^{0}\to \Lambda_{c}^{+}\pi^{-})/\mathcal{B}(\Xi_{c}^{0}\to \Xi^{-}\pi^{+})$ is measured to be  $0.38  \pm 0.04 \pm 0.04$. Combing with the world average value of $\BR(\Xi_{c}^{0}\to \Xi^{-}\pi^{+})$, the branching fraction $\BR(\Xi_{c}^{0}\to \Lambda_{c}^{+}\pi^{-})$ is deduced to be  $(0.54 \pm 0.05 \pm 0.05 \pm 0.12)\%$. Here, the uncertainties above are statistical, systematic, and from $\BR(\Xi_c^{0} \to \Xi^{-}\pi^{+})$, respectively.

\end{abstract}

\maketitle
\section{\boldmath Introduction}

The decay of charmed hadrons provides an ideal platform to study quantum chromodynamics (QCD). Usually, the charmed baryons decay via the transition of a $c$ quark into a $d$ or $s$ quark. However, baryons which contain both an $s$ and a $c$ quark also have a special class of decay, heavy-flavor-conserving nonleptonic decay, which proceeds via the decay of the $s$ quark. In such decays, the weak interaction among the light quarks can be well described by the short-distance effective Hamiltonian, since the emitted $\pi$ which has a low momentum due to the kinematic limit. Thus, the decay rate of the heavy-flavor-conserving nonleptonic decay process can be calculated by theory, and experimental measurements can be used to test the synthesis of heavy quark and chiral symmetries~\cite{Xic0_thy1,XiQ}.

\begin{figure}[htbp]
	\includegraphics[width=4.2cm]{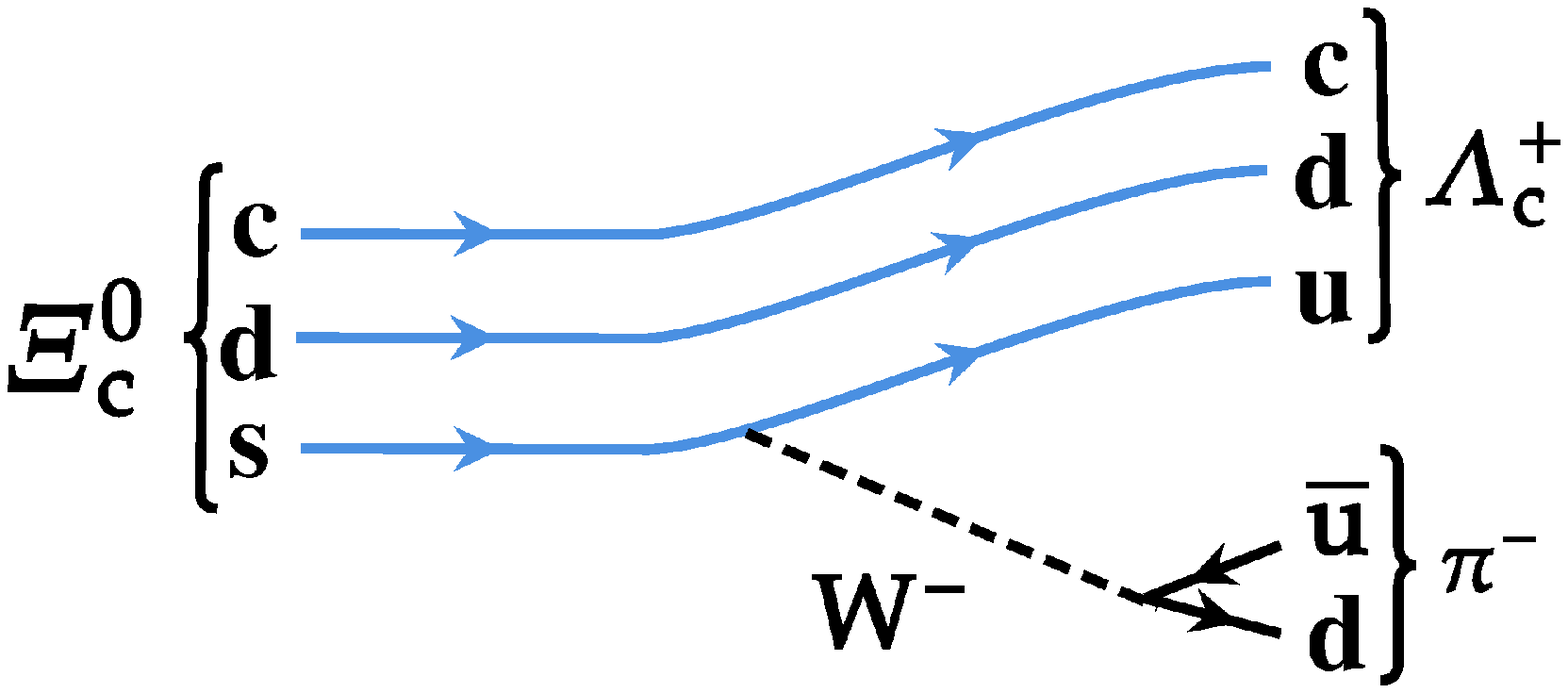}
    \includegraphics[width=4.2cm]{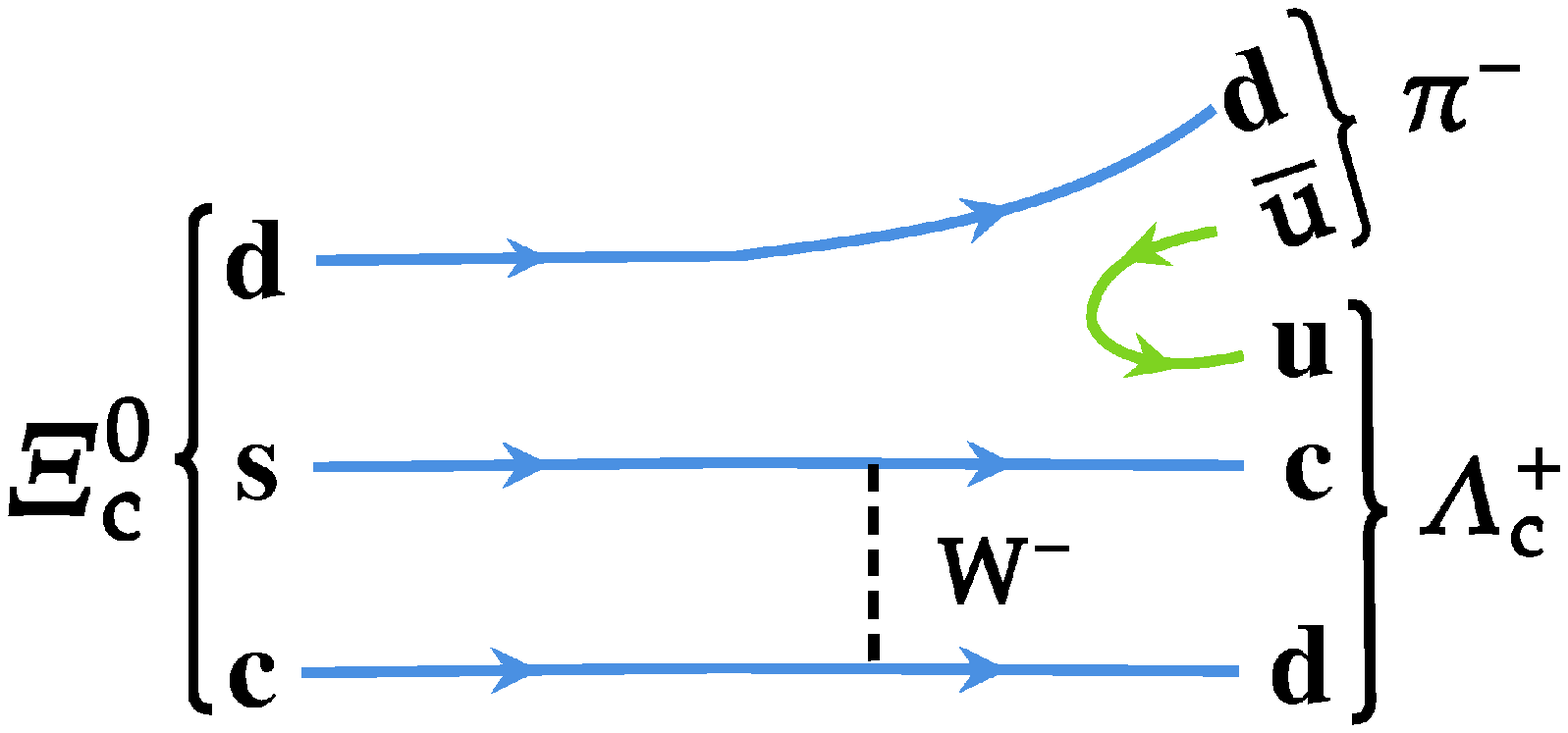}
    \put(-200,40){\bf (a)}
    \put(-80,40){\bf (b)}
	\caption{The Feynman diagrams of the (a) $s \to u(\bar{u}d)$ and (b) $cs \to dc$ modes of $\Xi_{c}^{0} \to \Lambda_{c}^{+} \pi^{-}$ decay.}\label{Feynman}
\end{figure}

The well-known $\Xi_{c}^{0}$ baryon consists of the $c$, $s$, and $d$ quarks and can decay via the disintegration of the $s$ quark with the $c$ quark acting as a spectator, i.e.\ $\Xi_{c}^{0} \to \Lambda_{c}^{+} \pi^{-}$. The decay width of $\Xi_{c}^{0} \to \Lambda_{c}^{+} \pi^{-}$ is based on the sizes of the $s$ quark decay amplitude of $s \to u(\bar{u}d)$ and the weak scattering amplitude $cs \to dc$, whose Feynman diagrams are shown in Fig~\ref{Feynman}.
Table~\ref{branching} summaries several previous theoretical predictions of the branching fraction of $\Xi_{c}^{0}\to \Lambda_{c}^{+}\pi^{-}$ using the measured $s \to u(\bar{u}d)$ amplitude and the weak scattering amplitude determined by the lifetimes of the $SU(3)$ anti-triplets $\Lambda_{c}^{+}$, $\Xi_{c}^{0}$, and $\Xi_c^{+}$~\cite{Xic0_theoryV,Xic0_theoryGR,Xic0_theoryFM,Xic0_theoryCH}. The large variation of these theoretical predictions is mainly due to different assumptions about the interference between the two strangeness-changing amplitudes.

\begin{table}[htbp]
	\footnotesize
	\begin{center}
		\caption{ Theoretical predictions on the branching fraction of $\Xi_{c}^{0} \to \Lambda_{c}^{+}\pi^{-}$ before experimental measurement ($10^{-3}$). All the results have been normalized using the current world average lifetimes of the $SU(3)$ anti-triplets~\cite{PDG,Xic0_theory2021}.}
		\vspace{0.2cm}
		\label{branching}
		\renewcommand\arraystretch{1.2}
		\setlength{\tabcolsep}{0.8mm}{
			\begin{tabular}{c  cc ccc}
				\hline\hline
				\tabincell{c}{(CLY)$^2$('92) \\ ~\cite{Xic0_thy1}} & \tabincell{c}{Voloshin\\ ~\cite{Xic0_theoryV}} &\tabincell{c}{Gronau \\ ~\cite{Xic0_theoryGR}}  &	 \tabincell{c}{Faller \\ ~\cite{Xic0_theoryFM}} &	   \tabincell{c}{(CLY)$^2$('06) \\ ~\cite{Xic0_theoryCH}}  & \\
				\hline
				0.39& $>(0.25\pm 0.15)$&$1.34\pm 0.53$ &$<3.9$ &0.17&\\
				\hline\hline
		\end{tabular}}
	\end{center}
\end{table}

The first experimental measurement on the branching fraction of $\Xi_{c}^{0}\to \Lambda_{c}^{+}\pi^{-}$ was performed by LHCb~\cite{Xic0}, finding a value ($0.55\pm0.02\pm0.18$)\%. The normalization of this result includes certain model-dependent assumptions based on heavy-quark symmetry and isospin. 

This result is closer to the prediction from Gronau and Rosner~\cite{Xic0_theoryGR} as listed in Table~\ref{branching}, which is calculated by assuming constructive interference between the two strangeness-changing amplitudes, meanwhile, the predicted branching fraction of $\Xi_{c}^{0}\to \Lambda_{c}^{+}\pi^{-}$ is less than 0.01\% for the destructive interference. Furthermore, the central value of LHCb result is generally larger than the theoretical predictions in Table~\ref{branching}. After the measurement from LHCb, $\BR(\Xi_{c}^{0} \to \Lambda_{c}^{+}\pi^{-})$ has been calculated to be $(0.58 \pm 0.21)\%$ from a study based on a constituent quark model~\cite{Niu:2021qcc}.

In this paper, we take $\Xi_{c}^{0}\to \Xi^{-}\pi^{+}$ decay as the reference mode and measure the branching fraction ratio of $\frac{\mathcal{B}(\Xi_{c}^{0}\to \Lambda_{c}^{+}\pi^{-})}{\mathcal{B}(\Xi_{c}^{0}\to \Xi^{-}\pi^{+})}$ via the $e^{+}e^{-}\to\Xi_{c}^{0}+anything$ inclusive decay process using Belle data samples.
The resulting $\BR(\Xi_{c}^{0}\to \Lambda_{c}^{+}\pi^{-})$ measurement, obtained using the world average value $\BR(\Xi_c^{0} \to \Xi^{-}\pi^{+})=(1.43\pm0.32)\%$~\cite{PDG,refemodeB}, is free of model dependent assumptions. Throughout this paper inclusion of charge-conjugate modes is implicitly assumed.

\section{\boldmath The data sample and the belle detector}

This analysis is based on data collected at or near the $\ones$,  $\twos$, $\threes$, $\fours$, and $\fives$ resonances by the Belle detector~\cite{detector,detector1} at the KEKB asymmetric-energy $e^+e^-$ collider~\cite{collider,collider1}. The data sample corresponds to an integrated luminosity of 983 fb$^{-1}$~\cite{detector1}. The Belle detector is a large solid-angle magnetic spectrometer that
consists of a silicon vertex detector, a 50-layer central drift chamber (CDC), an array of
aerogel threshold Cherenkov counters (ACC), a barrel-like arrangement of time-of-flight
scintillation counters (TOF), and an electromagnetic calorimeter comprised of CsI(Tl)
crystals (ECL) located inside a superconducting solenoid coil that provides a 1.5 T magnetic
field. An iron flux-return yoke instrumented with resistive plate chambers located outside
the coil is used to detect $K_{L}^{0}$ mesons and identify muons. A detailed description of the
Belle detector can be found in Refs.~\cite{detector,detector1}.

Signal Monte Carlo (MC) samples of one million events are generated with {\sc evtgen}~\cite{evtgen} to determine signal shapes and efficiencies for each $\Xi_{c}^{0}$ decay mode. The $e^{+}e^{-} \to c\bar{c}$ process is simulated using {\sc pythia}~\cite{PHTHIA}, and $\Xi_{c}^{0} \to \Lambda_{c}^{+}\pi^{-}$/$\Xi^{-}\pi^{+}$ decays are generated with a phase space model. The simulated events are processed with a detector simulation based on {\sc geant3}~\cite{geant}. Inclusive MC samples of $\Upsilon(1S$, $2S$, $3S)$ decays, $\fours \to B\bar{B}$, $\fives \to B_{(s)}^{(*)}{\bar{B}_{(s)}^{(*)}}$, and $e^{+}e^{-} \to q\bar{q}$ $(q=u,d,s,c)$ at  center-of-mass (C.M.) energies of 10.52, 10.58, and 10.867 GeV,
corresponding to 2 times the integrated luminosity of data, are used to optimize the signal selection criteria and to check possible peaking backgrounds.

\section{\boldmath Event selection criteria}

For well-reconstructed charged tracks in the signal mode, the impact parameters perpendicular to and along the beam direction with respect to the nominal interaction point  are required to be less than 1 cm and 3 cm, respectively. For the particle identification (PID) of a well-reconstructed charged track, information from different detector subsystems, including specific ionization in the CDC, time measurement in the TOF, and the response of the ACC, is combined to form a likelihood ${\mathcal L}_i$~\cite{pidcode} for particle species $i$, where $i$ = $p$, $\pi$, or $K$. Kaon candidates are required to have $\mathcal{L}_{K}/({\mathcal{L}}_{p}+{\mathcal{L}}_{K})$ $>$ 0.6 and $\mathcal{L}_{K}/({\mathcal{L}}_{K}+{\mathcal{L}}_{\pi})$ $>$ 0.6, with an approximately 89\% selection efficiency. For protons, the requirements are $\mathcal{L}_{p}/({\mathcal{L}}_{p}+{\mathcal{L}}_{\pi})$ $>$ 0.6 and  $\mathcal{L}_{p}/({\mathcal{L}}_{p}+{\mathcal{L}}_{K})$ $>$ 0.6, while for charged pions, the requirements are $\mathcal{L}_{\pi}/({\mathcal{L}}_{p}+{\mathcal{L}}_{\pi})$ $>$ 0.6 and $\mathcal{L}_{\pi}/({\mathcal{L}}_{K}+{\mathcal{L}}_{\pi})$ $>$ 0.6; these requirements are approximately 95\% efficient.

For $\Xi_{c}^{0}\to \Lambda_{c}^{+}\pi^{-}$, $\Lambda_{c}^{+}$ candidates are reconstructed via the $\Lambda_{c}^{+} \to$ $p$$K^{-}$$\pi^{+}$ decay mode and selected with $|M_{pK^-\pi^+}-m_{\Lambda_{c}^{+}}|$ $<$ 12 MeV/$c^2$ (within $\sim$ 3$\sigma$ of the nominal $\Lambda_{c}^{+}$ invariant mass, where $\sigma$ denotes the mass resolution). Hereinafter, $M_{x}$ represents the measured invariant mass and $m_{i}$ denotes the nominal mass of particle $i$~\cite{PDG}. The $\Lambda_{c}^{+}$ candidate is  combined with a $\pi^{-}$ to form the $\Xi_{c}^{0}$ candidate. To improve the momentum resolution and
suppress the backgrounds, vertex fits are performed for the selected $\Lambda_{c}^+$ and $\Xi_{c}^{0}$ candidates, and we require $\chi^{2}_{\rm vertex}/{\rm ndf} < 20$ with the corresponding efficiencies exceeding 90\%. To reduce combinatorial backgrounds, the scaled momentum $x_{p}= p^{*}/p_{\rm max}$ is required to be greater than 0.45. Here, $p^{*}$ is the momentum of $\Xi_{c}^{0}$ in the C.M.\ frame, and $p_{\rm max} =\sqrt{E_{\rm beam}^{2}-m^2_{\Xi_c^{0}}c^{4}}/c$, where $E_{\rm beam}$ is the beam energy in the $e^{+}e^{-}$ C.M. frame and $m_{\Xi_{c}^{0}}$ is the invariant mass of $\Xi_{c}^{0}$ candidates.


For the reference mode $\Xi_{c}^{0}\to \Xi^{-}\pi^{+}$, candidate $\Xi_{c}^{0}\to \Xi^{-}\pi^{+}$ events are selected using well-reconstructed tracks and PID in a way similar to the methods in Ref.~\cite{refemodeB}. The $\Lambda$ candidates are reconstructed in the decay $\Lambda \to p \pi^{-}$ with a production vertex significantly separated from the interaction point, and we define the $\Lambda$ signal region as $|M_{p\pi^{-}}-m_{\Lambda}|$ $<$ 3 MeV/$c^2$ ($\sim$ 2.5$\sigma$). The $\Xi^{-}$ candidate is reconstructed from the combination of selected $\Lambda$ and $\pi^{-}$ candidates. We define the $\Xi^{-}$ signal region as $|M_{\Lambda\pi^{-}}-m_{\Xi^{-}}|$ $<$ 6.5 MeV/$c^2$ ($\sim$ 3$\sigma$). Finally, the reconstructed $\Xi^{-}$ candidate is combined with a $\pi^{+}$ to form the $\Xi_{c}^{0}$ candidate. We perform vertex fits for the $\Lambda$, $\Xi^{-}$, and $\Xi_{c}^{0}$ candidates, and require $\chi^{2}_{\rm vertex}/{\rm ndf} < 20$. To suppress the combinatorial backgrounds, we require the flight directions of $\Lambda$ and $\Xi^{-}$ candidates, which are reconstructed from their fitted production and decay vertices, to be within five degrees of their momentum directions. The efficiency of this requirement is higher than 98\%. We also require the scaled momentum $x_{p} >$ 0.45. All the requirements on mass windows and scaled momenta above are optimized by maximizing $S/\sqrt{S+B}$, where $S$ is the  expected number of $\Xi_{c}^{0}$ events from signal MC samples using $\BR(\Xi_{c}^{0}\to \Lambda_{c}^{+}\pi^{-}) = 0.55\%$~\cite{Xic0} and $\BR(\Xi_{c}^{0} \to \Xi^{-} \pi^{+}) = 1.43\%$~\cite{PDG}, and $B$ is the number of expected background events in the $\Xi_c^{0}$ signal region from the inclusive MC samples.

\section{\boldmath Branching fraction of $\Xi_{c}^{0}\to \Lambda_{c}^{+}\pi^{-}$}

\begin{figure}[htbp]
	\includegraphics[width=7cm]{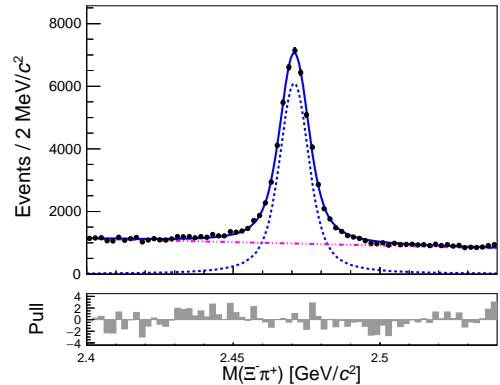}
	\caption{Fit to the invariant mass distribution of $\Xi^{-}\pi^{+}$ in data. The solid blue curve shows the best fitted result, the dashed blue curve shows the signal component, and the dot-dashed purple line shows the fitted backgrounds.}\label{refexic-data}
\end{figure}

After applying the above event selection criteria from the reference mode, the distribution of $M_{\Xi^{-}\pi^{+}}$ in the reference mode is shown in Fig.~\ref{refexic-data}. The yields of $\Xi_c^{0} \to \Xi^{-} \pi^{+}$ are extracted by an unbinned maximum-likelihood fit to the obtained $M_{\Xi^{-}\pi^{+}}$ distribution. The $\Xi_c^{0}$ signal shape is parameterized by a double-Gaussian function with the same mean value, and a first-order polynomial is used to describe the background shape. The central value of the signal function is fixed to the world average value~\cite{PDG}, and all other parameters in the fit are free to float. The fit result is shown in Fig.~\ref{refexic-data}, along with the pulls $(N_{\rm data}-N_{\rm fit})$/$\sigma_{\rm data}$, where the $\sigma_{\rm data}$ is the error on $N_{\rm data}$. The fitted  $\Xi_c^{0} \to \Xi^{-} \pi^{+}$ signal yield in data is $N_{\Xi^{-}\pi^{+}} $=$ (4.387\pm0.037)\times 10^{4}$.
The detection efficiency of $\Xi_c^{0} \to \Xi^{-} \pi^{+}$ is $16.4\%$ determined by fitting the corresponding $M_{\Xi^{-}\pi^{+}}$
spectrum from the signal MC sample, where efficiency correction factors due to PID have been included, and are discussed below.

\begin{figure}[htbp]
	\includegraphics[width=7cm]{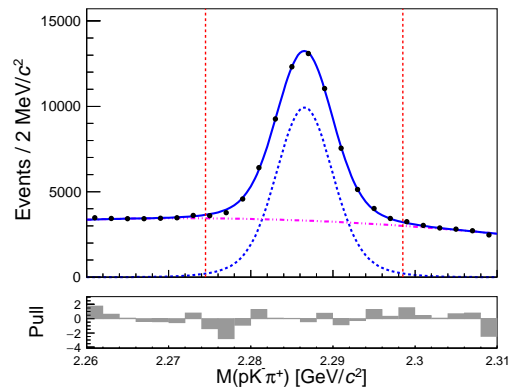}
	\caption{Fit to the invariant mass distribution of $\Lambda_{c}^{+}$ in data. The solid blue curve shows the best fitted result, the dashed blue curve shows the signal component, and the dot-dashed purple curve shows the fitted backgrounds. The red dashed lines show the defined $\Lambda_{c}^{+}$ signal region.}\label{dataLambdac}
\end{figure}

For the $\Xi_{c}^{0}\to \Lambda_{c}^{+}\pi^{-}$ signal mode, the invariant mass distribution of $\Lambda_{c}^{+}$ candidates is shown in Fig.~\ref{dataLambdac}. A double Gaussian function is used for the $\Lambda_{c}^{+}$ signal shape, while a second order polynomial is taken to describe the background shape. All the parameters in the fit are free. The $\Lambda_{c}^{+}$ mass window is indicated by the red dashed lines in Fig.~\ref{dataLambdac}. The fitted mass of $\Lambda_{c}^{+}$ is $(2286.55 \pm  0.03)$ MeV/$c^{2}$, which is consistent with the world average value~\cite{PDG}.

\begin{figure}[htbp]
	\includegraphics[width=7cm]{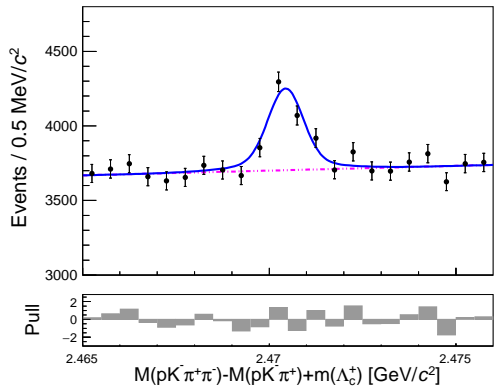}	
	\caption{The invariant mass distribution of $\Lambda_{c}^{+}\pi^{-}$ in data. The solid blue curve shows the best fitted result, the dot-dashed purple line shows the fitted backgrounds.
	}\label{dataxic0}
\end{figure}

After applying the mass window of $\Lambda_{c}^{+}$, the $M_{\Lambda_{c}^{+}\pi^{-}}$ spectrum is shown in Fig.~\ref{dataxic0},
together with the fit result and the corresponding pull distribution. The quantity $M_{p K^{-} \pi^+ \pi^-} - M_{p K^- \pi^+} + m_{\Lambda_c^+}$ is used, to remove the effect
of the $\Lambda_{c}^{+} \to p K^- \pi^+$ mass resolution. According to a study of the inclusive MC samples~\cite{topo}, previous Belle analysis~\cite{background2625}, and the $\Lambda_{c}^{+}$ sideband events, there is no peaking background in the 
$M_{\Lambda_{c}^{+}\pi^{-}}$ distribution in the range under study. Thus, the $\Xi_{c}^{0}$ signal shape is described by a double-Gaussian function, and a first-order
polynomial represents the backgrounds. The values of parameters in the double-Gaussian function are fixed to those obtained from the signal MC sample. The solid blue curve is the best fit result, and the dot-dashed purple line shows the fitted backgrounds.
The fitted $\Xi_{c}^{0}\to \Lambda_{c}^{+}\pi^{-}$ signal yield is $1467.7\pm134.5$. The statistical significance of the signal is 10.6$\sigma$. Here, the statistical significance is calculated using $\sqrt{-2\ln (\mathcal{L}_{0}/\mathcal{L}_{\rm max})}$, where $\mathcal{L}_{0}$ and $\mathcal{L}_{\rm max}$ are the maximized likelihoods without and with the signal component, respectively. The detection efficiency is found to be 14.6\% based on a fit to the $M_{\Lambda_{c}^{+}\pi^{-}}$ spectrum in the signal MC sample, where efficiency correction factors due to PID have been included, and are discussed below. The signal yields and detection efficiencies of  $\Xi_{c}^{0}\to \Lambda_{c}^{+}\pi^{-}$ and $\Xi_{c}^{0}\to \Xi^{-}\pi^+$ are summarized in Table~\ref{summary}.

\begin{table}[htbp]
	\caption{ Summary of the detection efficiencies ($\varepsilon$) and event yields ($N$) of $\Xi_{c}^{0} \to \Lambda_{c}^{+} \pi^{-}$ and  $\Xi_{c}^{0}\to \Xi^{-}\pi^{+}$. The uncertainties in the table are statistical only.}
	\vspace{0.2cm}
	\label{summary}
	\begin{tabular}{c c cc }
		\hline\hline
		& $\varepsilon$ &$N$ & \\
		\hline
		$\Xi_{c}^{0}\to \Xi^{-}\pi^{+}$&$16.4\%$&$43875 \pm 369$&\\
		$\Xi_{c}^{0}\to \Lambda_{c}^{+}\pi^{-}$&$14.6$\%& $1467\pm134$&\\
		\hline\hline
	\end{tabular}
\end{table}

The branching fraction ratio is calculated according to the formula,
\begin{footnotesize}
\begin{displaymath}
\begin{aligned}
\label{calculatBF}
		\frac{\mathcal{B}(\Xi_{c}^{0}\to \Lambda_{c}^{+}\pi^{-})}{\mathcal{B}(\Xi_{c}^{0}\to \Xi^{-}\pi^{+})}
	    & = \frac{ {N_{\Lambda_{c}\pi}} \times {\epsilon_{\Xi\pi}^{\rm ref}} \times \mathcal{B}(\Xi^{-}\to \Lambda \pi^{-}) \times \mathcal{B}(\Lambda\to p \pi^{-}) } {N_{\Xi\pi}\times\epsilon_{\Lambda_{c}\pi}^{\rm sig} \times \mathcal{B}(\Lambda_{c}^{+}\to pK^-\pi^{+})} \\
        &= 0.38 \pm 0.04({\rm stat.}) \pm 0.04 ({\rm syst.}),
\end{aligned}
\end{displaymath}
\end{footnotesize}
where $N_{\Lambda_{c}\pi}$ and $N_{\Xi\pi}$ are the signal yields of $\Xi_{c}^{0}\to \Lambda_{c}^{+}\pi^{-}$ and $\Xi_{c}^{0}\to \Xi^{-}\pi^{+}$  in data, respectively;
$\epsilon_{\Lambda_{c}\pi}^{\rm sig}$ and $\epsilon_{\Xi\pi}^{\rm ref}$ are the corresponding detection efficiencies; ${\mathcal{B}(\Lambda_{c}^{+}\to pK^-\pi^{+})}$, $\mathcal{B}(\Xi^{-} \to \Lambda \pi^{-})$, and $\mathcal{B}(\Lambda\to p \pi^{-})$ are the branching fractions taken from the Ref.~\cite{PDG}. Using the world average branching fraction $\BR(\Xi_c^{0} \to \Xi^{-}\pi^{+})=(1.43\pm0.32)\%$~\cite{refemodeB,PDG}, we measure $\mathcal{B}(\Xi_{c}^{0}\to \Lambda_{c}^{+}\pi^{-})=
(0.54 \pm 0.05 \pm 0.05 \pm 0.12)$\%, where the last uncertainty is from $\BR(\Xi_c^{0} \to \Xi^{-}\pi^{+})$.

\section{\boldmath Systematic Uncertainties}

There are several sources of systematic uncertainties for the measurement of the branching fraction of $\Xi_{c}^{0} \to \Lambda_{c}^{+} \pi^{-}$
as listed in Table~\ref{error1}, including detection-efficiency-related uncertainties, the branching fractions of intermediate states, as well as the fit method.

The detection-efficiency-related uncertainties include
those from tracking efficiency, PID efficiency, $\Lambda$ reconstruction efficiency, and the statistical uncertainty of the MC efficiency.
The tracking efficiency uncertainties cancel in the measured branching fraction ratio.
Using $D^{*+} \to D^{0}\pi^{+}$, $D^{0} \to K^{-}\pi^{+}$, and $\Lambda \to p\pi^{-}$ control samples,
the PID efficiency ratios of data and MC simulations are studied. For the signal decay of $\Xi_{c}^{0}\to \Lambda_{c}^{+}\pi^{-}$,
the PID efficiency ratios between the data and MC simulations are $\epsilon_{\rm data}/\epsilon_{\rm MC} = (100.0\pm0.9)\%,~(97.7\pm0.5)\%,~(99.8\pm1.0)\%$, and $(97.4\pm0.8)\%$ for the kaon, proton, pion from $\Xi_{c}^{0}$ decay, and the pion from $\Lambda_{c}^{+}$ decay, respectively. For the reference decay mode $\Xi_{c}^{0}\to \Xi^{-}\pi^{+}$, the PID efficiency ratios between the data and MC simulation are $\epsilon_{\rm data}/\epsilon_{\rm MC} =  (95.4\pm0.6)\%$ and $ (99.3\pm0.7)\%$ for the pion from $\Xi_{c}^{0}$ and the pion from $\Xi^{-}$, respectively. The central values of PID efficiency ratios are taken as the PID efficiency correction factors while their errors are taken as the systematic uncertainties due to PID for the selected tracks. 

Since the momentum distributions between signal mode and reference mode are different, the uncertainties on the PID efficiency for the pion do not completely cancel in the branching fraction ratio. When combining PID uncertainties, those for kaons and pions are added linearly, as they are taken from the same control sample: this procedure is conservative. The remaining uncertainties are added in quadrature, to yield the total PID systematic uncertainty on $\mathcal{B}(\Xi_{c}^{0}\to \Lambda_{c}^{+}\pi^{-})$, which is $4.0\%$. The uncertainty from $\Lambda$ reconstruction efficiency is 2.7\%, which is estimated based on its momentum distribution according to the previous study~\cite{Lambda_err}. We generate one million MC simulated events for both signal and reference decay modes, which introduce negligible systematic uncertainties (less than $0.3\%$) due to the statistical uncertainties of the detection efficiencies.

The uncertainties of branching fractions of $\Lambda_{c}^{+}\to pK^-\pi^+$, $\Xi^{-} \to \Lambda \pi^{-}$, and $\Lambda \to p\pi^{-}$ are $5.1\%$, $0.04\%$, and $0.8\%$, respectively~\cite{PDG}. They are added in quadrature to yield the total systematic uncertainty due to the branching fractions of intermediate states, which is $5.2\%$.

The systematic uncertainties from the fitting method include fit range, mass resolution, and the uncertainty in the $\Xi_c^0$ mass.
To consider the uncertainty due to mass resolution, we enlarge the mass resolution of the $\Xi_{c}^{0} \to \Lambda_{c}^{+}\pi^{-}$ signal shape by 10\% and take the difference in signal yields as the systematic uncertainty, which is $4.0\%$. The fit ranges are changed by 0.5 MeV/$c^{2}$ in both fits to $M_{\Lambda^{+}_{c}\pi^{-}}$ and $M_{\Xi^{-}\pi^{+}}$ spectra, and the deviations compared to the nominal fit results are taken as the systematic uncertainties, which are $4.4\%$ and $0.2\%$ for signal and reference modes, respectively.  In the fit to the $M_{\Lambda^{+}_{c}\pi^{-}}$ spectrum, the fitted $\Xi_{c}^{0}$ mass is $(2470.43 \pm 0.06)$ MeV/$c^{2}$ when we do not fix the central mass of signal function, which is consistent with the world average value~\cite{PDG} and the difference in signal yield compared to the nominal result is less than 0.1\%. Thus, the uncertainty from the $\Xi_{c}^{0}$ mass is neglected.

Assuming all the sources are independent and adding
them in quadrature, the total systematic uncertainty on
$\BR(\Xi_{c}^{0}\to \Lambda_{c}^{+}\pi^{-})$ is
obtained. All the systematic uncertainties are summarized
in Table~\ref{error1}, where the uncertainty of 22.4\% on $\BR(\Xi_{c}^{0}\to \Xi^{-}\pi^+)$~\cite{PDG}
is not included and treated as an independent systematic uncertainty.

\begin{table}[htbp]
	\caption{ The systematic uncertainties on the measurement of $\BR(\Xi_{c}^{0}\to \Lambda_{c}^{+}\pi^{-})$. The uncertainty of $\BR(\Xi_{c}^{0}\to \Xi^{-}\pi^+)$ is taken as an independent uncertainty and not listed in this table. }
	\vspace{0.2cm}
	\label{error1}
	\begin{tabular}{c c  }
		\hline\hline
		Sources  & Value (\%) \\
		\hline
		PID efficiency&4.0\\
		$\Lambda$ selection & 2.7\\
		Branching fractions of intermediate states&5.2\\
		Mass resolution &4.0\\
	    Fit range &4.4\\
	    MC statistical&0.3\\
		Total &9.3\\
		\hline\hline
	\end{tabular}
\end{table}

\section{\boldmath conclusion}
In summary, using the entire data sample of 983 fb$^{-1}$ collected with the Belle detector, we perform a model independent measurement on the branching fraction of $\Xi_{c}^{0}\to \Lambda_{c}^{+}\pi^{-}$. The branching fraction ratio is calculated to be $$\frac{\mathcal{B}(\Xi_{c}^{0}\to \Lambda_{c}^{+}\pi^{-})}{\mathcal{B}(\Xi_{c}^{0}\to \Xi^{-}\pi^{+})}= 0.38 \pm 0.04\pm 0.04. $$ Taking $\BR(\Xi_c^{0} \to \Xi^{-}\pi^{+})=(1.43\pm0.32)\%$~\cite{PDG}, the absolute branching fraction of $\Xi_{c}^{0}\to \Lambda_{c}^{+}\pi^{-}$ is measured to be $(0.54 \pm 0.05 \pm 0.05 \pm 0.12)\%$, where the uncertainties are statistical, systematic, and from $\BR(\Xi_c^{0} \to \Xi^{-}\pi^{+})$, respectively. This result is consistent with the measurement by
LHCb~\cite{Xic0}, and although less precise than their model-dependent result. It is larger than the theoretical predictions~\cite{Xic0_theoryV,Xic0_theoryGR,Xic0_theoryFM,Xic0_theoryCH}. This result, once combined with the improved $\BR(\Xi_c^{0} \to \Xi^{-}\pi^{+})$ expected from Belle II, can constrain theoretical models more stringently.

\section{\boldmath ACKNOWLEDGMENTS}
This work, based on data collected using the Belle detector, which was
operated until June 2010, was supported by 
the Ministry of Education, Culture, Sports, Science, and
Technology (MEXT) of Japan, the Japan Society for the 
Promotion of Science (JSPS), and the Tau-Lepton Physics 
Research Center of Nagoya University; 
the Australian Research Council including grants
DP180102629, 
DP170102389, 
DP170102204, 
DE220100462, 
DP150103061, 
FT130100303; 
Austrian Federal Ministry of Education, Science and Research (FWF) and
FWF Austrian Science Fund No.~P~31361-N36;
the National Natural Science Foundation of China under Contracts
No.~11675166,  
No.~11705209;  
No.~11975076;  
No.~12135005;  
No.~12175041;  
No.~12161141008; 
Key Research Program of Frontier Sciences, Chinese Academy of Sciences (CAS), Grant No.~QYZDJ-SSW-SLH011; 
the Ministry of Education, Youth and Sports of the Czech
Republic under Contract No.~LTT17020;
the Czech Science Foundation Grant No. 22-18469S;
Horizon 2020 ERC Advanced Grant No.~884719 and ERC Starting Grant No.~947006 ``InterLeptons'' (European Union);
the Carl Zeiss Foundation, the Deutsche Forschungsgemeinschaft, the
Excellence Cluster Universe, and the VolkswagenStiftung;
the Department of Atomic Energy (Project Identification No. RTI 4002) and the Department of Science and Technology of India; 
the Istituto Nazionale di Fisica Nucleare of Italy; 
National Research Foundation (NRF) of Korea Grant
Nos.~2016R1\-D1A1B\-02012900, 2018R1\-A2B\-3003643,
2018R1\-A6A1A\-06024970, RS\-2022\-00197659,
2019R1\-I1A3A\-01058933, 2021R1\-A6A1A\-03043957,
2021R1\-F1A\-1060423, 2021R1\-F1A\-1064008, 2022R1\-A2C\-1003993;
Radiation Science Research Institute, Foreign Large-size Research Facility Application Supporting project, the Global Science Experimental Data Hub Center of the Korea Institute of Science and Technology Information and KREONET/GLORIAD;
the Polish Ministry of Science and Higher Education and 
the National Science Center;
the Ministry of Science and Higher Education of the Russian Federation, Agreement 14.W03.31.0026, 
and the HSE University Basic Research Program, Moscow; 
University of Tabuk research grants
S-1440-0321, S-0256-1438, and S-0280-1439 (Saudi Arabia);
the Slovenian Research Agency Grant Nos. J1-9124 and P1-0135;
Ikerbasque, Basque Foundation for Science, Spain;
the Swiss National Science Foundation; 
the Ministry of Education and the Ministry of Science and Technology of Taiwan;
and the United States Department of Energy and the National Science Foundation.
These acknowledgements are not to be interpreted as an endorsement of any
statement made by any of our institutes, funding agencies, governments, or
their representatives.
We thank the KEKB group for the excellent operation of the
accelerator; the KEK cryogenics group for the efficient
operation of the solenoid; and the KEK computer group and the Pacific Northwest National
Laboratory (PNNL) Environmental Molecular Sciences Laboratory (EMSL)
computing group for strong computing support; and the National
Institute of Informatics, and Science Information NETwork 6 (SINET6) for
valuable network support.



\begin{thebibliography}{**}

\bibitem{Xic0_thy1} H. Y. Cheng, C. Y. Cheung, G. L. Lin, Y. C. Lin, T. M. Yan, and H. L. Yu, Phys. Rev. D \textbf{46}, 5060 (1992).
\bibitem{XiQ} M. B. Voloshin, Phys. Lett. B \textbf{476}, 297 (2000).	
\bibitem{Xic0_theoryV}  M. B. Voloshin, Phys. Rev. D {\bf 100}, 114030 (2019).
\bibitem{Xic0_theoryGR} M. Gronau and J. L Rosner, Phys. Lett. B {\bf 757}, 330 (2016).
\bibitem{Xic0_theoryFM} S. Faller and T. Mannel, Phys. Lett. B {\bf 750}, 653 (2015).
\bibitem{Xic0_theoryCH} H. Y. Cheng, C. Y. Cheung, G. L. Lin, Y. C. Lin, T. M. Yan, and H. L. Yu, JHEP {\bf03}, 028 (2016).

\bibitem{Xic0} R. Aaij {\it et al.} (LHCb Collaboration), Phys. Rev. D {\bf 102}, 071101 (2020).
\bibitem{PDG} P.A. Zyla {\it et al.} (Particle Data Group), Prog. Theor. Exp. Phys. {\bf 2020}, 083C01 (2020) and 2021 update.
\bibitem{Xic0_theory2021} H. Y. Cheng, Charmed Baryon Physics Circa 2021, [arXiv:2109.01216].
\bibitem{Niu:2021qcc} P.~Y.~Niu, Q.~Wang, and Q.~Zhao, Phys. Lett. B \textbf{826}, 136916 (2022).
\bibitem{refemodeB} Y. B. Li {\it et al.} (Belle Collaboration), Phys. Rev. Lett. {\bf 122}, 082001 (2019).
\bibitem{detector} A. Abashian {\it et al.} (Belle Collaboration), Nucl. Instrum. Methods Phys. Res., Sect. A {\bf 479}, 117(2002).
\bibitem{detector1} J. Brodzicka {\it et al.}, Prog. Theor. Exp. Phys. {\bf 2012}, 04D001 (2012).
\bibitem{collider} S. Kurokawa and E.~Kikutani, Nucl. Instrum. Methods Phys. Res., Sect. A {\bf 499}, 1 (2003).
\bibitem{collider1} T. Abe {\it et al.}, Prog. Theor. Exp. Phys. {\bf 2013}, 03A001 (2013).

\bibitem{evtgen} D. J. Lange, Nucl. Instrum. Methods Phys. Res., Sect. A {\bf 462}, 152 (2001).
\bibitem{PHTHIA} T. Sjöstrand, S. Mrenna, and P. Z. Skands,  JHEP {\bf 05}, 026 (2006).
\bibitem{geant} R. Brun {\it et al.}, CERN Report No. DD/EE/84-1, 1984.
\bibitem{pidcode} E. Nakano, Nucl. Instrum. Methods Phys. Res., Sect. A {\bf 494}, 402 (2002).
\bibitem{topo}  X. Y. Zhou, S. X. Du, G. Li, and C. P. Shen, Comput. Phys. Commun. {\bf 258}, 107540 (2021).
\bibitem{background2625} S. -H. Lee {\it et al.} (Belle Collaboration), Phys. Rev. D. {\bf 89}, 091102(R) (2014).
\bibitem{Lambda_err} Y. Kato {\it et al.} (Belle Collaboration), Phys. Rev. D \textbf{94}, 032002 (2016).

\end{thebibliography}
\end{document}